\documentclass[a4paper,
               keeplastbox,   
               hyphens,      
               ]{jacow}
\usepackage{pdfpages,multirow,ragged2e} %
\makeatletter%
	\ifboolexpr{bool{xetex}}
	 {\renewcommand{\Gin@extensions}{.pdf,%
	                    .png,.jpg,.bmp,.pict,.tif,.psd,.mac,.sga,.tga,.gif,%
	                    .eps,.ps,%
	                    }}{}
\makeatother

\ifboolexpr{bool{xetex} or bool{luatex}} 
 {}                                      

\usepackage[USenglish]{babel}

\usepackage{caption, subcaption}
\usepackage{lipsum}
\usepackage{float}
\usepackage{hyperref}
\usepackage{IEEEtrantools}
\usepackage{tabularx}
\usepackage{tabulary}

\begin{document}

\title{Upgrade of the fast analogue intra-pulse phase feedback at SPARC\_LAB}

\author{L. Piersanti\thanks{luca.piersanti@lnf.infn.it}, M. Bellaveglia, A. Gallo, R. Magnanimi, S. Quaglia, M. Scampati, G. Scarselletta, \\B. Serenellini, S. Tocci
\\ INFN - Laboratori Nazionali di Frascati, Via Enrico Fermi 54, 00044 Frascati (RM), Italy}
	
\maketitle
\begin{abstract}
 SPARC\_LAB is a facility designed for the production of FEL radiation and the exploration of advanced acceleration techniques using a high brightness electron photo-injector. Specifically, particle-driven plasma wakefield acceleration (PWFA) necessitates exceptional beam stability, in order to minimize the jitter between the driver and witness beams. This requirement directly translates into RF phase jitter minimization, since a velocity bunching (RF compression) working point is employed at SPARC\_LAB for acceleration.  In the past, a fast intra-pulse phase feedback system has been developed to stabilize the klystron RF pulse. This allowed to reach a phase stability of S-band power units (both driven by PFN modulators) below \SI{50}{\femto s} rms. However, in order to meet the more stringent requirements of PWFA scheme, some upgrades of this feedback system have been recently carried out. A prototype has been tested on a C-band klystron driven by a solid-state modulator, in order to investigate the possibility for an additional improvement resulting from the inherently more stable power source. In this paper the preliminary measurement results obtained at SPARC\_LAB after such upgrades will be reviewed.
\end{abstract}

\section{SPARC\_LAB facility at LNF}
SPARC\_LAB facility\cite{Ferrario_sparc} was born as an R\&D activity to develop a high brightness e- photo-injector for SASE-FEL experiments. 
The installation of the machine at LNF began in fall  2004. The first tests on the RF gun and measurement of beam parameters started in 2006.
From that moment on a relentless R\&D program has been carried out performing many experiments, such as: 

\begin{itemize}
\item{SASE and seeded FEL;}
\item{Thomson back-scattering;}
\item{THz generation;}
\item{Plasma focusing and acceleration (Particle WakeField Acceleration - PWFA).}
\end{itemize}

Nowadays SPARC\_LAB is a multidisciplinary experimental facility that combines:
\begin{itemize}
\item{high brightness S-band photoinjector;}
\item{conventional booster linac, made of two S-band travelling wave accelerating structure (SLAC-type) and one C-band constant impedance structure;}
\item{plasma acceleration/focusing stage for PWFA (together with a brand new plasma laboratory for capillary R\&D);}
\item{12 m undulator for FEL applications (both SASE and seeded);}
\item{\SI{200}{\tera W} (\SI{5}{J}, \SI{25}{\femto s}, \SI{10}{Hz} rep. rate) class laser (FLAME) for LWFA and experiments of laser interaction with matter.}
\end{itemize}

In the last years the R\&D activities have been devoted to plasma focusing and acceleration in view of the future LNF flagship project EuPRAXIA\@SPARC\_LAB\cite{Ferrario_eupraxia}, that aims at building the first plasma driven FEL user facility, with a compact linac fully realized with the X-band technology. 

The first activity that has been pursued has been the setup of a test-bench for plasma characterization and density measurement, which is of vital importance to correctly match the beam parameters for efficient acceleration \cite{Biagioni}. Then, active and passive plasma lens experiments were carried out, in order to exploit the strong forces generated within the plasma capillary to focus the beam \cite{Pompili_active, Pompili_active2}, or exploiting the plasma wakefields to manipulate the longitudinal phase space improving the peak brightness \cite{Shpakov}.
All this work led to a much deeper understanding of the plasma physics and finally to the first plasma acceleration experiments. The first proof of plasma acceleration has been obtained not aiming to the maximum gradient, but instead to the energy spread minimization \cite{Pompili_accel}, in order to preserve as much as possible the quality of the beam. Then, during the November 2022 experimental run, the maximum accelerating gradient has been recorded to be $\approx$1 GV/m.
Moreover, also FEL experiments (both SASE and seeded) have been carried out \cite{Pompili_sase,Galletti_seeded}, in order to demonstrate the capability and the quality of a plasma accelerated beam to generate FEL radiation. This proof-of-principle experiments represent a fundamental milestone in the use of plasma-based accelerators, contributing to the development of next-generation compact facilities for user-oriented applications, such as the future flagship project of LNF EuPRAXIA\@SPARC\_LAB. 

\section{SPARC\_LAB RF system}
\subsection{General layout}
\begin{figure*}[hbt]
   \centering
   \includegraphics*[width=1.5\columnwidth]{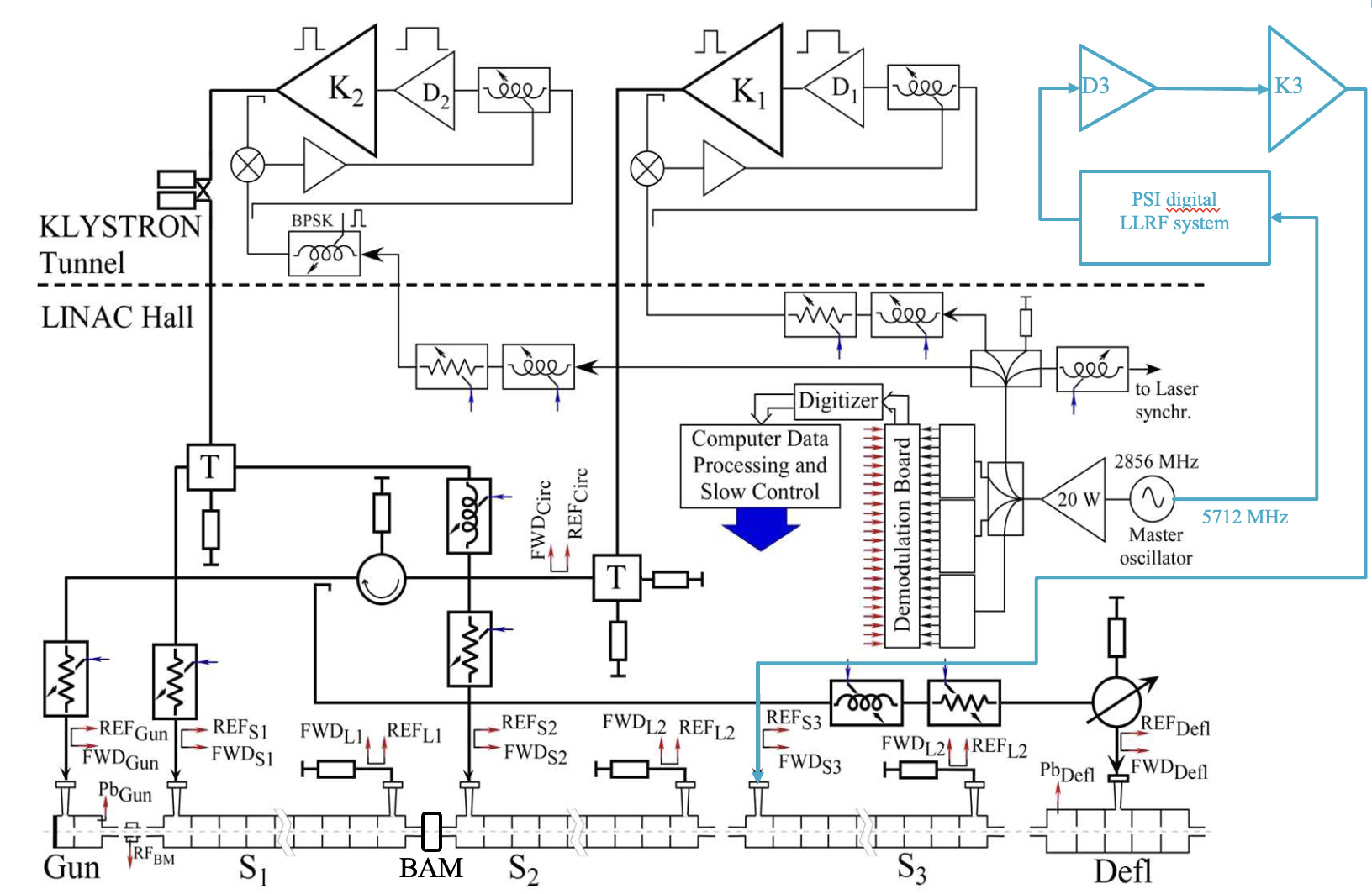}
   \caption{SPARC\_LAB RF system layout}
   \label{fig:RF_layout}
\end{figure*}

A sketch of SPARC\_LAB current RF system is reported in Fig.\ref{fig:RF_layout}.
The Reference Master Oscillator (RMO) is a custom ultra low phase noise oven controlled crystal oscillator made by Laurin A.G. that provides coherent outputs at \SI{2856}{MHz}, \SI{5712}{MHz} and \SI{2142}{MHz}. Both the S-band and C-band references are amplified and distributed to the RF power plants and low level RF (LLRF) front and back-end. The \SI{2856}{MHz} is also used to phase-lock the photo-cathode laser and the FLAME laser. The \SI{2142}{MHz} is used to downconvert to baseband the signal from a Beam Arrival Monitor (BAM) cavity used for beam timing measurement.

The two S-band RF power plants are based on PFN modulators (from Puls-PlasmaTechnik - PPT) and \SI{45}{MW} pulsed klystrons from Thales (TH2128C) driven by \SI{300}{W} solid state amplifiers from Microwave Amplifiers.
The C-band plant combines a solid state modulator from ScandiNova and a \SI{50}{MW} klystron from Canon (E37202) with a similar driver amplifier.

The first S-band klystron drives the RF gun and the deflecting cavity used for beam diagnostics. In the waveguide network there is a circulator in the common path and a variable attenuator, phase shifter and RF switch in the deflector line. All these components are in SF6 atmosphere.
The second S-band klystron serves the two SLAC-type travelling wave accelerating structures and its RF pulse is compressed with a SLED type pulse compressor.
The C-band plant is dedicated to the constant impedance high gradient (\SI{35}{MV/m}) structure used as energy booster.

\subsection{LLRF systems}

The S-band LLRF system has been designed and realized in 2006 by the RF group of the LNF. It employs commercial RF components for signal manipulation and ADCs for signal acquisition. The front-end is a 24 channel direct conversion system based on custom Pulsar Microwave I/Q mixers. The base-band I/Q signals are subsequently sampled by commercial ADC cards (National Instruments 5105, 60 MHz, 12 bit). The analog back-end employs connectorized RF components (trombone and electronic phase shifters for coarse and fine tuning respectively, variable attenuators and Binary Phase Shift Keying for pulse compressor phase modulation). Being a fully analog system, it has no pulse shaping capabilities. The noise of the front-end has been estimated to limit the phase readout to a floor of the order of \SI{50}{\femto s}.
The slow feedbacks against drifts of RF field amplitude and phase are performed via control system. 

The C-band LLRF, instead, is a digital system designed and realized by PSI in the framework of TIARA collaboration. It is made of a 16 channels front-end (>\SI{80}{dB} isolation between channels) that down-converts the RF signals to an IF of \SI{39.667}{MHz} before digitization (16 bit ADC). The analog bandwidth of the front-end is larger than \SI{30}{MHz}. The system has also pulse shaping capabilities but only signal detection. No feedback can be performed directly from the LLRF, but  also in this case the slow feedbacks are implemented in the control system. The phase error is <$\pm$\SI{0.05}{deg} and <\SI{0.1}{\%} for the amplitude. The back-end has a differential I/Q vector modulator with a bandwidth >\SI{40}{MHz} and an added jitter <\SI{10}{\femto s}.

Recently a regional funding of \SI{6.1}{Meuro} has been awarded to the LNF for the consolidation of SPARC\_LAB facility (SABINA project). The LLRF system will be fully upgraded with digital systems provided by Instrumentation Technologies (Libera LLRF). These are temperature stabilized, FPGA based digital LLRF with a low noise front-end that demodulates the RF down to \SI{44.625}{MHz} before digital conversion (\SI{14}{bit}, \SI{119}{MHz}, \SI{5}{MHz} BW). The system has the capability to perform pulse-to-pulse amplitude and phase feedback on 2 independent chnannels, and the vector modulator can geerate an arbitrary pulse shape from a mask given by the user (\SI{16}{bit} DAC, \SI{15}{MHz} BW). This upgrade will overcome some known limitations of the actual system, mostly concerning the front-end noise floor, the temperature stabilization (to minimize drifts) and the possibility to have an arbitrary pulse shape.

\section{Fast intra-pulse phase feedback}
One of the main contributions to RF phase jitter in a pulsed electron linac like SPARC\_LAB is due to the high voltage jitter of the modulator that feeds the klystron. High voltage fluctuations are, in fact, directly converted to phase jitter by the klystron tube, and, especially for pulse forming network (PFN) modulators, this contribution can be of the order of hundreds of \SI{}{\femto s}. Solid state modulators, instead, are natively more stable (the state of the art of HV pulse to pulse jitter can be as low as \SI{10}{ppm} (even though this value is highly dependent on the particular "modulator specimen" as studied at PSI for their RF power plants \cite{geng_PSI}), and for typical applications rarely require a further stabilization. 

\subsection{First feedback design}
\begin{figure}[hbt]
   \centering
   \includegraphics*[width=\columnwidth]{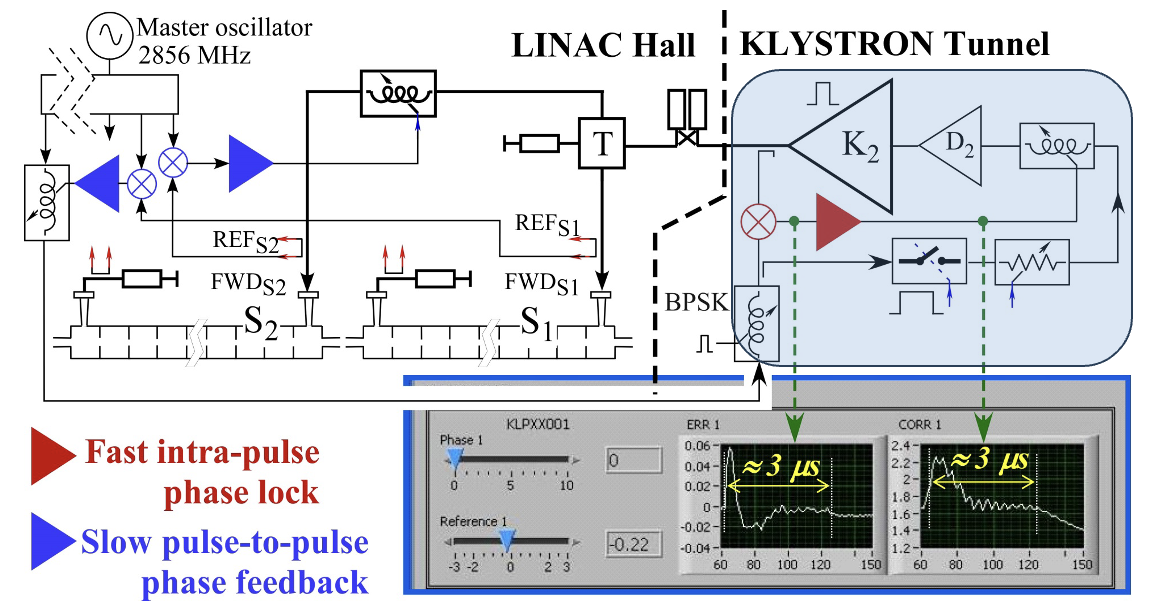}
   \caption{SPARC\_LAB klystron n.2 RF system. Highlighted the block diagram of the fast intra-pulse feedback.}
   \label{fig:K2loop}
\end{figure}

The S-band modulators at SPARC\_LAB have been acquired in 2006 and are made with PFN technology. While for the first experiments conducted this was not a limitation, as the development of the machine progressed and increasingly challenging experiments were proposed, a reduction in RF phase jitter became necessary. 
For this reason in 2008 the first version of a fast intra-pulse phase feedback has been designed, realized and successfully tested at SPARC\_LAB.
The block diagram of the feedback system (also referred to as "klystron loop") can be seen from Fig.\ref{fig:RF_layout} and is highlighted in Fig.\ref{fig:K2loop} for the klystron number 2.

\begin{figure}[hbt]
   \centering
   \includegraphics*[width=\columnwidth]{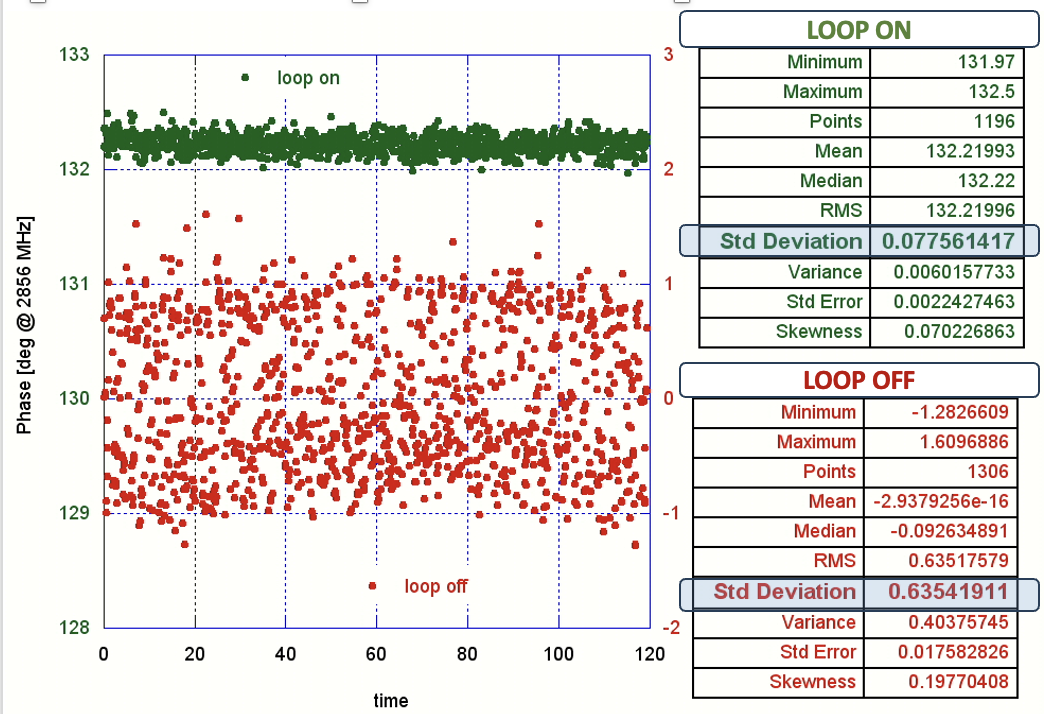}
   \caption{Test at SPARC\_LAB of the fast intra-pulse feedback. In red the RF phase jitter of Klystron forward signal without the feedback, in green with feedback on. The jitter reduction is almost one order of magnitude.}
   \label{fig:v1results}
\end{figure}

The actuator is a fast phase shifter (Pulsar Microwave ST-G9-411), that has a \SI{3}{dB} bandwidth >\SI{10}{MHz}, and the error amplifier is based on a two stage current feedback operational amplifiers (CLC410). On the bottom of Fig.\ref{fig:K2loop} the typical error signals and correction signals are also shown. The feedback response is pretty fast, after $\approx$ \SI{1}{\micro s} the output settles to the nominal value. After a thorough laboratory characterization, the system has been successfully tested at SPARC\_LAB, and the results are shown in Fig.\ref{fig:v1results}. RF pulses acquired for \SI{120}{s} have been analyzed by the LLRF system averaging the phase over a \SI{100}{\nano s} time window. The measured jitter reduction was substantial, from \SI{0.64}{deg} rms to \SI{0.077}{deg} rms. 
Recent measurements on the RF power plants, performed after PFN modulator upgrade with a Libera LLRF system, show an even lower jitter of \SI{0.041}{deg}.

\subsection{System upgrade}
The very good RF stability reached with the first version of the klystron loop is however still not enough for plasma acceleration. In fact, if a stable and reproducible acceleration of driver and witness beam is required (especially in the case of EuPRAXIA@SPARC\_LAB where external users will perform experiments with FEL radiation) a further step has to be made. Beam dynamics simulations, with the so-called velocity bunching RF compression scheme used at SPARC\_LAB, pose an upper limit to the maximum jitter of the RF stations to \SI{10}{\femto s}.
One way to tackle the problem could be replacing the old PFN modulators with new solid state ones. In this way the starting amount of jitter can be drastically reduced. This approach has however some practical drawbacks: it is very expensive, it requires a non negligible amount of time from tender to the receipt of the components.
For this reason an upgrade of the feedback electronics of the existent plants has been planned instead. Moreover this upgrade will be realized on the C-band station, in order to test it on the only solid state modulator available at SPARC\_LAB. This will give us an idea of the jitter limit that could be reached with such approach. Since the C-band line at SPARC\_LAB is only used as energy booster, it is possible to test the performance of the prototype during the experimental run with minimal interference.

The upgrade involved: (i) the phase shifter used as actuator (that in the first version has an insertion loss strongly dependent on the control voltage and in this case has to work at \SI{5712}{\mega Hz}), (ii) the operational amplifiers used in the error amp circuit (that have been chosen with a constant GBW product of $\approx$ \SI{200}{\mega Hz} and (iii) the presence of an internal slow feedback ($G_{slow}/G_{fast} \approx 10^{-3}$) that should keep the phase value close to the desired value even when the RF pulse is off. The goal of such upgrade is to reach a steady state within \SI{100}{\nano s}, minimizing the phase modulation required at the beginning of the RF pulse.

\section{Preliminary measurement results}

\begin{figure}[hbt]
   \centering
   \includegraphics*[width=0.8\columnwidth]{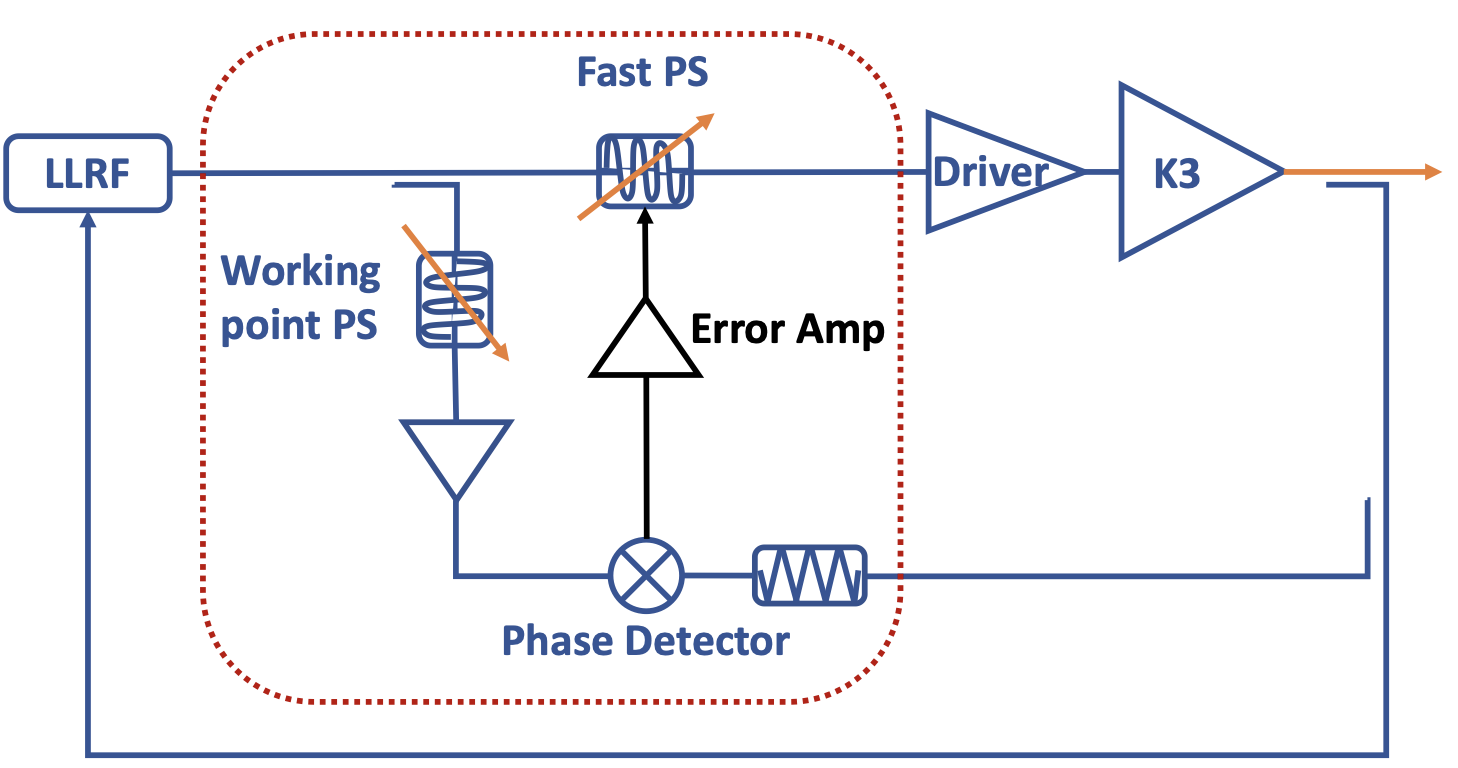}
   \caption{Block diagram of the experimental setup during the tests on C-band line at SPARC\_LAB of the new intra-pulse feedback.}
   \label{fig:exp_setup}
\end{figure}

In order to verify the functionality of the new klystron loop, detailed laboratory tests have been carried out. 
After succeeding in this preliminary phase, the new feedback system will be installed at SPARC\_LAB and RF measurements will be carried out. A block diagram of the experimental setup is shown in Fig.\ref{fig:exp_setup}.

Due to a late restart of the machine after summer shutdown and some hardware faults that delayed the laboratory tests of the klystron loop electronics, we have not tested at SPARC\_LAB the new error amplifier yet. A measurement campaign has been carried out with a prototype (without the low gain feedback functionality) and the results are reported in Tab.\ref{tab:meas_results}. A reduction on the C-band klystron  phase jitter of a factor 4 (from \SI{0.12}{deg} to \SI{0.031}{deg}, corresponding to \SI{14.9}{\femto s}) has been observed. 

\begin{table}[!htb]
    \centering
    \caption{Preliminary measurement results obtained at SPARC\_LAB with error amplifier prototype}
    \label{tab:meas_results}	
	\begin{tabulary}{\columnwidth}{LCC}
		\hline
		Signal: Klystron 3 FWD & Jitter (deg) & Jitter (fs)\\
		\hline
        loop OFF & 0.12 & 58.4\\
        \hline
        loop ON & 0.031 & 14.9\\

		\end{tabulary}
\end{table}
These are very promising results, even though there are still some open points that we would like to investigate further in the future, such as: (i) reducing the group delay of the signals placing the loop electronics close to the klystron, (ii) choose a larger bandwidth phase shifter, (iii) quantify and minimize the effect of EM and grounding noise in the klystron gallery. Moreover the capabilities of the slow loop, that keeps the phase constant also during the dead time between two consecutive RF pulses, still need to be tested.

\section{SUMMARY}

In this paper the design, the tests on bench and the preliminary results of a simpler prototype of the new fast intra-pulse phase feedback at SPARC\_LAB have been reported. Encouraging results have been observed, that allowed to reduce the RF phase jitter on the C-band klystron forward signal from \SI{0.12}{deg} rms to \SI{0.031}{deg} rms. There are still open points to be addressed in order to optimize the experimental setup in the future, for instance: placing the electronics close to the klystron to reduce the delay of cables, choose another phase shifter with larger bandwidth, take care of EM and grounding noise in the klystron gallery. Moreover the capabilities of the slow loop, that keeps the phase constant also during the dead time between two consecutive RF pulses, still need to be tested. Given the importance of this activity, a long R\&D program to optimize and deploy such feedback system also to S-band klystron is foreseen in the next months at LNF.

\bibliography{main}

\begin{thebibliography}{10}
\providecommand{\url}[1]{#1}
\csname url@samestyle\endcsname
\providecommand{\newblock}{\relax}
\providecommand{\bibinfo}[2]{#2}
\providecommand{\BIBentrySTDinterwordspacing}{\spaceskip=0pt\relax}
\providecommand{\BIBentryALTinterwordstretchfactor}{4}
\providecommand{\BIBentryALTinterwordspacing}{\spaceskip=\fontdimen2\font plus
\BIBentryALTinterwordstretchfactor\fontdimen3\font minus
  \fontdimen4\font\relax}
\providecommand{\BIBforeignlanguage}[2]{{%
\expandafter\ifx\csname l@#1\endcsname\relax
\typeout{** WARNING: IEEEtran.bst: No hyphenation pattern has been}%
\typeout{** loaded for the language `#1'. Using the pattern for}%
\typeout{** the default language instead.}%
\else
\language=\csname l@#1\endcsname
\fi
#2}}
\providecommand{\BIBdecl}{\relax}
\BIBdecl
\renewcommand{\BIBentryALTinterwordstretchfactor}{4}

\bibitem{Ferrario_sparc}
Ferrario \emph{et~al.}, ``{SPARC\_LAB} present and future,'' \emph{Nuclear
  Instruments and Methods in Physics Research Section B: Beam Interactions with
  Materials and Atoms}, vol. 309, pp. 183--188, 2013.

\bibitem{Ferrario_eupraxia}
Ferrario \emph{et~al.}, ``{EuPRAXIA@SPARC\_LAB Design study towards a compact
  FEL facility at LNF},'' \emph{Nuclear Instruments and Methods in Physics
  Research Section A: Accelerators, Spectrometers, Detectors and Associated
  Equipment}, vol. 909, pp. 134--138, 2018.

\bibitem{Biagioni}
Biagioni \emph{et~al.}, ``Electron density measurement in gas discharge plasmas
  by optical and acoustic methods,'' \emph{Journal of Instrumentation},
  vol.~11, no.~08, p. C08003, aug 2016.

\bibitem{Pompili_active}
Pompili \emph{et~al.}, ``Focusing of high-brightness electron beams with
  active-plasma lenses,'' \emph{Phys. Rev. Lett.}, vol. 121, p. 174801, 2018.

\bibitem{Pompili_active2}
Pompili \emph{et~al.}, ``Experimental characterization of active plasma lensing
  for electron beams,'' \emph{Applied Physics Letters}, vol. 110, p. 104101,
  2017.

\bibitem{Shpakov}
Shpakov \emph{et~al.}, ``Longitudinal phase-space manipulation with beam-driven
  plasma wakefields,'' \emph{Physical review letters}, vol. 122, no.~11, p.
  114801, 2019.

\bibitem{Pompili_accel}
Pompili \emph{et~al.}, ``Energy spread minimization in a beam-driven plasma
  wakefield accelerator,'' \emph{Nature Physics}, vol.~17, no.~4, pp. 499--503,
  2021.

\bibitem{Pompili_sase}
Pompili \emph{et~al.}, ``Free-electron lasing with compact beam-driven plasma
  wakefield accelerator,'' \emph{Nature}, no. 7911, pp. 659--662, 2022.

\bibitem{Galletti_seeded}
Galletti \emph{et~al.}, ``Stable operation of a free-electron laser driven by a
  plasma accelerator,'' \emph{Physical Review Letters}, vol. 129, p. 234801,
  2022.

\bibitem{geng_PSI}
Geng \emph{et~al.}, ``{RF Jitter and Electron Beam Stability in the SwissFEL
  Linac},'' in \emph{Proc. FEL'19}, ser. Free Electron Laser Conference,
  no.~39.\hskip 1em plus 0.5em minus 0.4em\relax JACoW Publishing, Geneva,
  Switzerland, nov 2019, paper WEP037, pp. 400--403.

\end{thebibliography}

%
%
	
		
		
	
%
\end{document}